\documentclass{azh}

\usepackage{graphicx}
\usepackage{natbib}
\usepackage{lscape}

\def\degr{\hbox{$^\circ$}}

\def\arcsec{\hbox{$^{\prime\prime}$}}

\begin{document} 

\title{Regular Chains of Star Formation Regions in Spiral Arms and Rings 
       of Disk Galaxies}
       
\author{A.~S.~Gusev}

\institute{Sternberg Astronomical Institute, Lomonosov Moscow State University, 
                Moscow, Russia}

\date{Received December 30, 2022; revised March 27, 2023; accepted 
      March 27, 2023}
\offprints{Alexander~S.~Gusev, \email{gusev@sai.msu.ru}}

\titlerunning{Regular Chains of Star Formation Regions}
\authorrunning{Gusev}

\abstract{The regularity in the distribution of young stellar groups along 
the spiral arms of galaxies, first discovered by Bruce and Debra Elmegreen 
in 1983, was considered a rather rare phenomenon. However, recent studies 
of the spatial regularities in the distribution of the young stellar populations 
along the arms of the spiral galaxies NGC~628, NGC~895, NGC~4321, NGC~5474, 
NGC~6946, as well as along the rings of the spiral galaxy NGC~6217 and the 
lenticular galaxy NGC~4324, have revealed that this spatial (quasi) regularity 
and/or the presence of regular chains of star-forming regions is a fairly common 
phenomenon. Across all galaxies, the characteristic regularity scale is 
350–500~pc or a multiple thereof. It should be noted that theoretical models 
predict an instability scale of a stellar-gas disk on the order of a few kpc, 
which is several times larger than what has been observed. The paper is partly 
based on the report presented at the Modern Stellar Astronomy 2022 Conference 
held at the Caucasian Mountain Observatory of the Sternberg Astronomical 
Institute, Moscow State University, on November 8–10, 2022. \\


{\bf DOI:} 10.1134/S1063772923050049 \\
}

\maketitle

\section{INTRODUCTION}

In 1983, B.G.~Elmegreen and D.M.~Elmegreen \citep{elmegreen1983} brought 
attention to the fact that neighboring HII regions in the spiral arms of 
some galaxies are located at equal distances from each other. 
\cite{elmegreen1983} noted the rarity of this phenomenon: among 200 
galaxies exhibiting well-defined spiral structures and numerous 
star-forming regions, visually detectable chains of HII regions were only 
present in 22 of them. Among these, demonstrated regularity in the 
distribution of HII regions within a single spiral arm, while only 7 galaxies 
exhibited regularity in two arms. The characteristic distances between 
adjacent HII regions varied from 1 to 4~kpc in different galaxies 
\citep{elmegreen1983}.

Later, \cite{efremov2009,efremov2010} discovered a similar regularity in the 
distribution of HI clouds in the spiral arms of our own Galaxy. He estimated 
the average distance between HI superclouds in the Carina spiral arm to be 
1.5~kpc, and in the Cygnus arm, 1.3~kpc \citep{efremov2009}. Using ultraviolet 
images obtained through the Galaxy Evolution Explorer (GALEX) program, he 
also identified a regular chain of stellar complexes within the north-western 
arm of the Andromeda Nebula with a characteristic distance of 1.1~kpc. 
\cite{efremov2010} attributed this regularity to the presence of a consistent 
magnetic field in that arm.

The rarity of spatial regularities in the distribution of stellar complexes 
or HII regions in the spiral arms of galaxies appears to be quite 
understandable. In the simplest case of gravitational instability within a 
gaseous medium, the characteristic distance (derived in \cite{elmegreen1983} 
directly from the \cite{safronov1960} equation) between adjacent regions of 
star formation along the spiral arms is
\begin{equation}
\label{equation:l1}
\Lambda = 2c_{\rm g}^2/G\Sigma_{\rm g},
\end{equation}
where $c_{\rm g}$ is the speed of sound in a gas, $\Sigma_{\rm g}$ is the 
surface density of the gas, and $G$ is the gravitational constant. Note that 
the characteristic scale $\Lambda$ formally does not depend on the disk 
stability parameter $Q$ (Toomre parameter) in this case; however, parameter 
$Q$ defines the characteristic time $\tau$ for the development of 
instabilities in a rotating disk:
\begin{eqnarray}
\tau \sim (1-Q^2)^{-1/2}. \nonumber
\end{eqnarray}

When the magnetic field is considered, Eq.~(\ref{equation:l1}) takes 
the form
\begin{eqnarray}
\Lambda = 2c_{\rm g}^2/G\Sigma_{\rm g}\lambda_{\rm mag}, \nonumber
\end{eqnarray}
where coefficient $\lambda_{\rm mag}$ depends on $Q$ and the magnetic 
field energy density $w$ in a complex way (see \cite{elmegreen1983}, 
Eqs.~(11)–(15) and \cite{lynden1966}, equations). The magnetic field 
allows the development of instabilities in the gaseous disk at $Q>1$ 
\citep{elmegreen1983,lynden1966}.

Coefficient $\lambda_{\rm mag}=1-1.5$ for the entire range of observed 
parameters $Q$ and $w$; in particular, for $Q=1$, 
$\lambda_{\rm mag}=1.1-1.2$ (see \cite{elmegreen1983}, Table~3).

In a more general case, theoretical studies of the gravitational instability 
of stellar-gas \citep{jog1984a,jog1984b,romeo2013} and multicomponent disks 
\citep{rafikov2001} have revealed that the perturbation wavelength 
$\Lambda = 2\pi/k$ ($k$ is the wavenumber of the instability) depends on 
a comprehensive set of parameters: $\Sigma_{\rm g}$, $c_{\rm g}$, surface 
density of the stellar disk $\Sigma_{\rm s}$, the dispersion of stellar 
velocities $\sigma_{\rm s}$, and the shape of the disk's rotation curve 
$v(r)$ (refer to corresponding equations in \cite{leroy2008,marchuk2018}). 
In this case, the wavelength of perturbations in a two-component medium 
corresponds to the minimum of the dispersion curve for two components 
(fragmentation in the medium occurs at the wavelength with the fastest 
growth). The regularity in the distribution of star-forming regions and 
their precursor molecular clouds necessitates the constancy of the 
aforementioned physical parameters of the stellar-gas disk across 
sufficiently large scales within the galactic disk and a wide range of 
galactocentric distances $r$. However, such constancy should not often 
occur in classical galactic disks.

In addition to theoretical models associated with the gravitational 
instability of a stellar-gas disk, there are alternative models that 
describe the fragmentation of gas filaments \citep{inutsuka1997,mattern2018}, 
single- and multi-component spiral arms \citep{elmegreen1994b,inoue2018}, 
and rings \citep{elmegreen1994a} in galaxy disks, attributed to the growth 
of azimuthal perturbations along the arms (rings). All these models also 
indicate the dependence of the perturbation wavelength on the same 
parameters of the interstellar medium, the stellar disk, and the magnetic 
field as the models of the gravitational (magnetogravitational) instability 
of the disk.

It is worth noting that all theoretical calculations predict the perturbation 
wavelength (or the characteristic distance between neighboring star-forming 
regions) on the order of several kiloparsecs, considering typical values of 
the parameters of the stellar and interstellar medium 
\citep{elmegreen1983,marchuk2018,inoue2021}.

Over the past decade, numerous studies have emerged examining the regularity 
in the distribution of the young stellar populations and molecular clouds 
along the spiral arms and rings of galaxies. These studies have reshaped our 
understanding of the prevalence of this phenomenon and its characteristic 
spatial scales. On the other hand, in the past 10–15 years, new infrared and 
radio observational data on the stellar component, HI and H$_2$ in galaxies 
have become available. The angular resolution of these data allows for the 
analysis of spatial parameters of the stellar disk and gas on subkiloparsec 
scales for nearby galaxies (projects BIMA SONG \citep{helfer2003}, THINGS 
\citep{walter2008}, HERACLES \citep{leroy2009}, CARMA \citep{rahman2012}, 
SINGS \citep{kennicutt2003}). This enables direct measurements of the scale 
of regularity within theoretical models.

The objective of this study is to summarize the data regarding regularities 
in the distribution of young stellar populations and molecular clouds along 
the spiral arms and rings of galaxies of various types, These findings, 
obtained by our own research team as well as other scientific groups, will 
be compared with calculations based on theoretical models.

\section{RESULTS OF STUDIES ON REGULARITIES IN THE DISTRIBUTION OF YOUNG 
         STELLAR POPULATION}
         
\begin{figure*}
\centerline{\includegraphics[width=17.8cm]{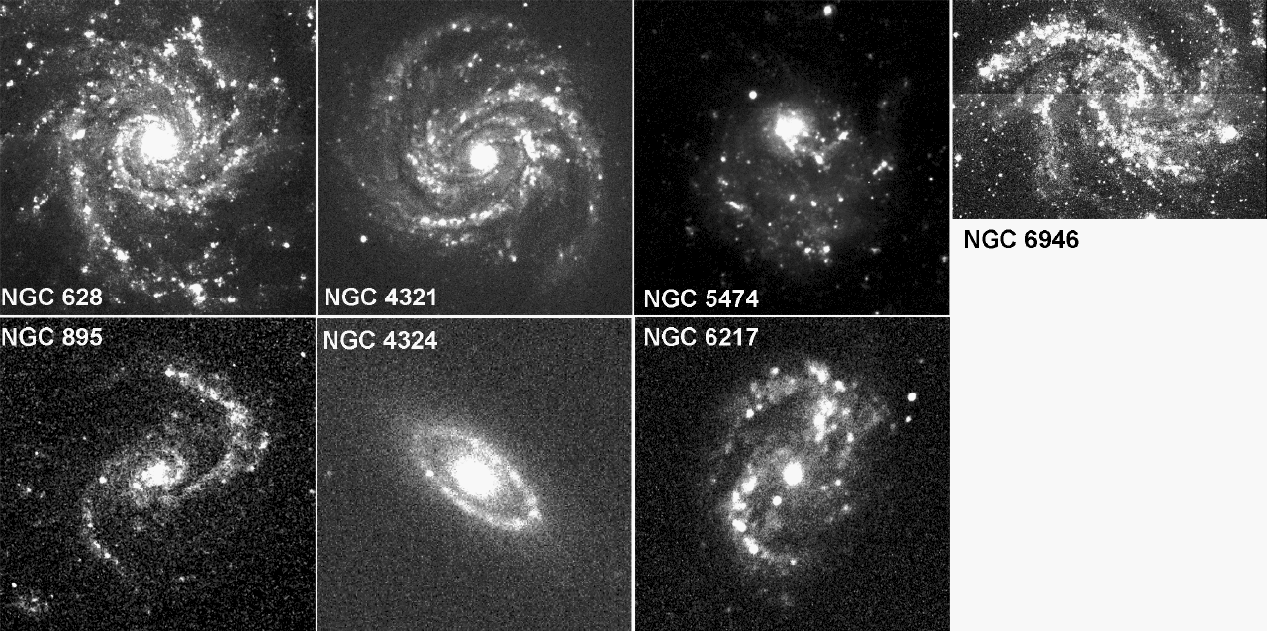}}
\caption{Images of galaxies with regular chains of star-forming regions 
found in their spiral arms or rings. The image of NGC~4324 is given in the 
$u$ band; other galaxies, in the $U$ band. North is up, east is to the left.
\label{figure:fig1}}
\end{figure*}

\cite{gusev2013} studied the distribution of star-forming regions 
of various sizes and luminosities (star clusters, complexes, and HII regions) 
in the grand design galaxy Sc NGC~628 (M74; 
Fig.~\ref{figure:fig1}\footnote{The image of NGC~4321 in the figure obtained 
in \cite{marcum2001} was taken from the {\sc ned} database; NGC~4324, from the 
SDSS survey database (http://www.sdss.org/dr13/) \citep{albareti2017}; 
images of other galaxies are taken from the observations according to the 
author's programs.}), which is observed at a close distance of 7.2~Mpc 
according to \cite{sharina1996}. The galaxy is observed almost ''face-on'' 
(inclination $i=7\degr$ according to \cite{sakhibov2004} and other sources).

\begin{figure}
\vspace{12mm}
\centerline{\includegraphics[width=7cm]{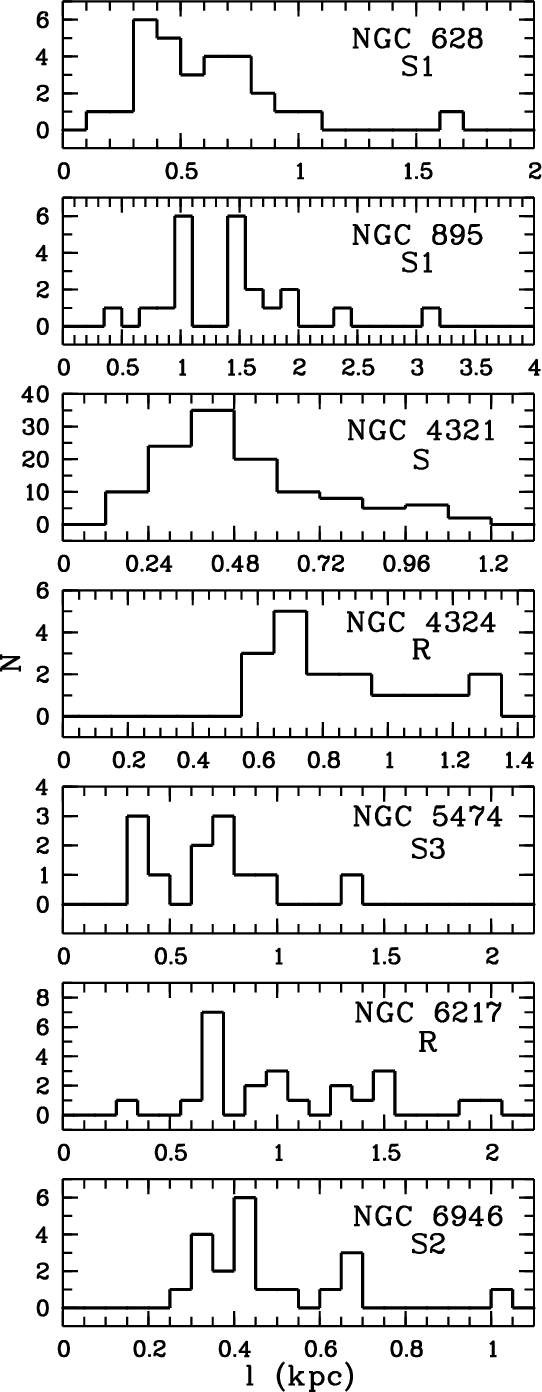}}
\caption{Examples of distance distributions between neighboring 
star-forming regions (local brightness maxima) in spiral arms and 
galactic rings.
\label{figure:fig2}}
\end{figure}

\begin{figure}
\vspace{5mm}
\centerline{\includegraphics[width=8cm]{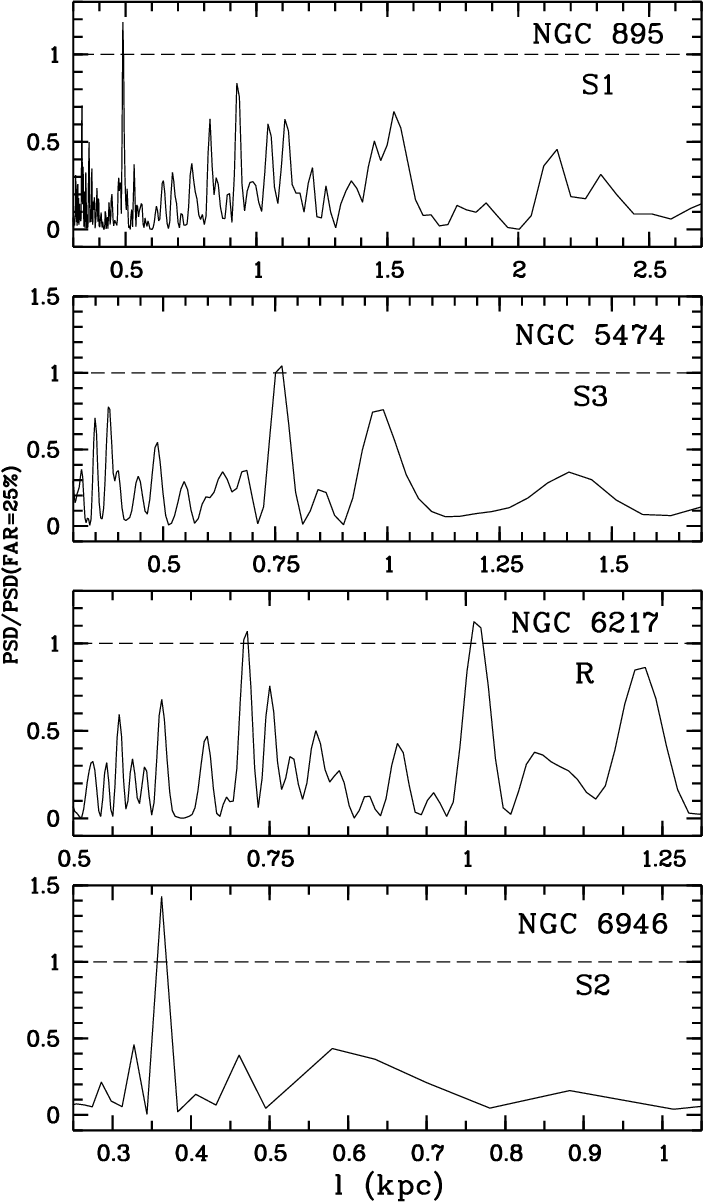}}
\caption{Examples of normalized power spectral density of the 
distribution function of local brightness maxima in photometric bands 
sensitive to the presence of the young stellar population in spiral 
arms and rings of galaxies. The PSD values $>1$ correspond to the 
probability of a false detection of the selected frequency (wavelength) 
$<25\%$.
\label{figure:fig3}}
\end{figure}

In contrast to previous studies that relied on visual identification of 
regular chains of star-forming regions, our research employed a more 
mathematically rigorous algorithm. Initially, if required, the galaxy was 
oriented to the ''face-on'' position. The spiral arms were approximated with 
logarithmic spirals. The next step involved aperture photometry along the 
spiral using elliptical apertures with large eccentricities. The aperture 
centers were positioned along the spiral at specific intervals, and the 
semimajor axes of the ellipses were chosen to cover the entire width of the 
spiral arm, perpendicular to the tangent at the aperture center. Subsequently, 
brightness profiles were constructed in bands sensitive to the presence 
of young stellar population (ultraviolet FUV, NUV, $UB$, and H$\alpha$ in 
optics, as well as 8~$\mu$m in the IR range). The maxima of the profiles 
and the distances $\lambda$ between the maxima were determined. Further 
analysis was carried out using histograms of $\lambda$ distributions 
(Fig.~\ref{figure:fig2}) and Fourier analysis (Fig.~\ref{figure:fig3}). 
In later studies, we utilized the Fourier analysis-based methods developed 
in \cite{scargle1982,horne1986,press1989,vp2018} to analyze time series with 
a high-noise signal.

The probability of false additional peaks on the periodograms increases with 
a decrease in the number of objects and a decrease in the signal-to-noise 
ratio \citep{vp2018}. Most of the galaxies examined in this section had more 
than 10 regions of concentration of young stellar population. With a smaller 
number of objects, we can only discuss local regular chains.

The numerical simulation carried out in \cite{vp2018} demonstrated that when 
the number of objects $\le10$, the resulting peak with the maximum power 
density always corresponds to the actual wavelength (frequency). Any false 
peaks, if present, exhibit lower power spectral density (see \cite{vp2018}, 
Fig.~25).

The studies of \cite{gusev2013} in FUV, $U$, and H$\alpha$ have shown that 
the young stellar population in both spiral arms of NGC~628 exhibits 
a regular distribution with a characteristic scale $\Lambda\approx400$~pc, 
which is four times smaller than the estimates in \cite{elmegreen1983} for 
this galaxy. At the same time, large stellar complexes observed only in one 
of the arms of NGC~628 are indeed located at a distance of $4\Lambda=1.6$~kpc 
from each other, while samples of smaller stellar groups (clusters and 
associations) are located at characteristic distances $2\Lambda=800$~pc from 
each other in both spiral arms \citep{gusev2013}.

It was also shown in \cite{gusev2013} that the presence or absence of shock 
waves does not affect the regularity in the distribution of the young stellar 
population. Signs of shock waves are only observed in one of the arms of 
NGC~628, where large stellar complexes are absent. However, a regular 
distribution of star-forming regions persists along both arms with the same 
characteristic distance.

\cite{elmegreen2018} investigated the distribution of IR dust emission 
sources in the 8~$\mu$m band, which represent regions of modern star 
formation with an age $<1-2$~Myr, in the slightly inclined ($i=24\degr$ 
according to the {\sc leda} 
database\footnote{http://leda.univ-lyon1.fr/}) and relatively close 
(distance 16.2~Mpc according to the {\sc ned} 
database\footnote{http://ned.ipac.caltech.edu/}) SABb-SABbc-type 
galaxy NGC~4321 (M100). This well-known galaxy has two spiral arms with 
numerous filaments (Fig.~\ref{figure:fig1}).

\cite{elmegreen2018} identified 27 regular chains of dust condensations 
in the galaxy, consisting of a total of 147 sources. The characteristic 
distance between neighboring objects $\lambda\approx410$~pc, which aligns 
with the previously obtained value for the star-forming regions in NGC~628.

\cite{gusev2020} studied the distribution of star-forming regions in the 
ring of the barred galaxy NGC~6217 (type SBb; Fig.~\ref{figure:fig1}), for 
which the regularity of distances between neighboring regions 
($\approx1.3$~kpc) was suspected in \cite{artamonov1999}. Using 
observational data in NUV, $U$, and H$\alpha$, the authors confirmed the 
regularity in the distribution of the young stellar population in the ring 
with a characteristic distance $\Lambda\approx700$~pc. The presence of 
the $3\Lambda/2\approx1.0$~kpc mode (Figs.~\ref{figure:fig2}, 
\ref{figure:fig3}), also observed in the ring of NGC~6217, may indicate 
the presence of the characteristic distance $\Lambda/2$, which is close 
to the value of 400~pc obtained for NGC~628 \citep{gusev2013} and 
NGC~4321 \citep{elmegreen2018}. It should be noted that NGC~6217 is located 
three times farther from us ($d=20.6$~Mpc) than NGC~628, which poses 
a challenge in analyzing the spatial distribution in NGC~6217 at scales 
on the order of 500~pc or less \citep{gusev2020}.

It should be noted that the galaxy ring is located near the corotation 
radius, where the presence of shock waves is not expected \citep{gusev2020}.

In 2022, two independent research teams published papers investigating the 
regularity in the distribution of the young stellar population in spiral 
arms and rings of four galaxies.

The first study \citep{gusev2022} considered the distribution of the young 
stellar galaxy in the spiral arms of the Scd galaxies NGC~895, NGC~5474, 
and NGC~6946. The analysis was performed in the FUV, $U$, and H$\alpha$ 
bands, with additional observations at 8~$\mu$m for NGC~6946. Despite 
the same morphological type, these galaxies differ greatly in structure. 
NGC~895 is a typical two-arm grand design galaxy; NGC~5474 is an 
asymmetric galaxy with three arms on one side of the center; the 
four-armed NGC~6946 appears to be flocculent, although its spirals are 
of a wave nature \citep{kendall2011,ghosh2016} (Fig.~\ref{figure:fig1}).

In NGC~895, the most distant galaxy ($d=32.7$~Mpc) among those considered 
in this section, \cite{elmegreen1983} found a regular chain of six stellar 
complexes with a characteristic distance of 1.4~kpc in only one of the 
two spiral arms. The analysis carried out in \cite{gusev2022} revealed 
spatial regularities in the distribution of the young stellar population 
in both spiral arms. The first arm confirmed the $1.4-1.5$~kpc scale 
obtained in \cite{elmegreen1983}, and a characteristic distance 
$\approx1.0$~kpc was found. The Fourier analysis also indicated the 
presence of the $\Lambda\approx500$~pc scale (Figs.~\ref{figure:fig2}, 
\ref{figure:fig3}). In the second spiral arm, the characteristic distance 
scales of $\approx1.0$, $\approx1.5$, and $\approx2.0$~kpc were found, 
with the latter being confidently confirmed by the Fourier analysis 
as well.

In the short inner arm of NGC~5474, \cite{gusev2022} found a chain of 
five star-forming regions forming a regular sequence with $\Lambda=430$~pc 
and $2\Lambda=860$~pc. In the middle arm of the galaxy, six star-forming 
regions constituted a regular chain with characteristic distances between 
the neighbors of $\approx660$~pc and $\approx1.15$~kpc. This chain was 
previously noted in \cite{elmegreen1983}. Two characteristic distances 
were also found in the far spiral arm: $\approx390$~pc and the doubled 
$\approx740$~pc. The latter was also confirmed by the Fourier analysis 
data, which gave a value of 750~pc (Figs.~\ref{figure:fig2}, 
\ref{figure:fig3}).

In NGC~6946, regular chains of star-forming regions were not visually 
apparent in any of the spiral arms. However, \cite{gusev2022} revealed 
a spatial regularity in the distribution of the young stellar 
population and/or the presence of regular chains of star formation regions 
in all spiral arms. A regular chain of five star-forming regions with 
a characteristic distance $\approx400$~pc was found in one of the inner 
arms of the galaxy. The second of the inner spiral arms exhibited 
a regularity in the distribution of the young stellar population in 
the UV and optics (distance 360–380~pc), as well as IR 
(distances $\approx310$, $\approx550$, and $\approx890$~pc). The minimum 
scales $\approx310$ and 360–380~pc were confirmed by the Fourier analysis. 
In the third, outer spiral of NGC~6946, regular distributions of the young 
stellar population of $\sim300$~pc in the optical and 420–440~pc in the 
IR were revealed according to \cite{gusev2022}. The fourth spiral arm 
displayed a regular distribution with a characteristic distance 
$\Lambda\approx400$~pc in the optical, UV, and IR sources.

In the second paper of 2022, \cite{proshina2022} discovered 
a regularity in the distribution of young stellar complexes and 
HII~regions in the ring of the lenticular (!) SA0+ galaxy NGC~4324 
(Fig.~\ref{figure:fig1}). By studying this galaxy in the FUV, NUV, 
$u$ bands and H$\alpha$ line, the authors found that the distances between 
neighboring HII regions within the ring were $\Lambda\approx650$~pc, while 
the stellar complexes visible in $u$ had distances of $\approx1.0$~kpc. 
At the same time, relatively younger star formation regions (HII~regions 
with an age of up to 10 Myr) often occupied positions in the middle between 
relatively old ($t\sim100$~Myr) stellar complexes.

\begin{table}
\begin{center}
\caption{Characteristic distances $\lambda$ between neighboring 
star-forming regions in spiral arms and galactic rings.}
\label{table:tab1}
\begin{tabular}{|r|l|l|}
\hline\hline
NGC & S/R$^a$ & $\lambda$, kpc \\
\hline
 628 & S1      & 0.4, 0.8, 1.6 \\
     & S2      & 0.4, 0.8 \\
 895 & S1      & 0.5, 1.0, 1.5 \\
     & S2      & 1.0, 1.5, 2.0 \\
4321 & S (IR)  & 0.41 \\
4324 & R       & 0.65, 1.0 \\
5474 & S1      & 0.43, 0.86 \\
     & S2      & 0.66, 1.15 \\
     & S3      & 0.39, 0.75 \\
6217 & R       & 0.7, 1.0 \\
6946 & S1      & 0.4 \\
     & S2      & 0.37 \\
     & S2 (IR) & 0.31, 0.55, 0.89 \\
     & S3      & 0.3 \\
     & S3 (IR) & 0.43 \\
     & S4      & 0.4 \\
     & S4 (IR) & 0.4 \\[1mm]
\hline
\end{tabular}
\end{center}
$^a$ S1, ... is the number of the spiral arm, R is the ring of 
the galaxy \\
\end{table}

The numerical results obtained in \cite{proshina2022} closely agree with 
the data from \cite{gusev2020} concerning the ring of NGC~6217. It should 
be noted that the galaxy NGC~4324 is even further away from us than 
NGC~6217 ($d=26.2$~Mpc \citep{tully2013}), and investigating potetial 
regularities on smaller scales (300-500~pc) does not appear possible.

Table~\ref{table:tab1} provides the summary data on the characteristic 
distances between neighboring star-formation regions in the spiral arms 
and rings of the seven galaxies studied in recent years.

The results demonstrate that the characteristic regularity scale $\Lambda$ 
in all these galaxies is 350–500~pc or a multiple of thereof.

\section{PERTURBATION WAVELENGTH $\Lambda$ IN THE GAS AND STELLAR-GAS 
         DISK OF THE GALAXY NGC~628}
         
\begin{figure*}
\centerline{\includegraphics[width=17.8cm]{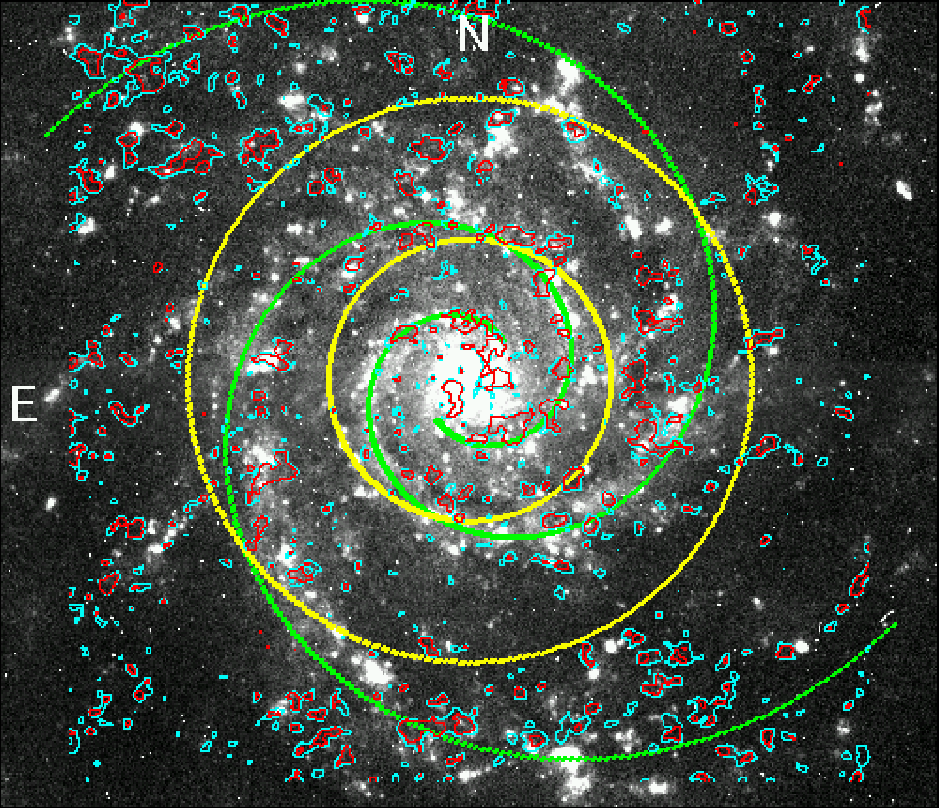}}
\caption{Image of NGC~628 in the $U$ band. The green curves are 
logarithmic spiral arms. The yellow circles have radii $60\arcsec$ and 
$120\arcsec$. Red and blue colors mark the isolines of $\Lambda=600$~pc 
and $\Lambda=800$~pc, respectively.
\label{figure:fig4}}
\end{figure*}

Using the BIMA SONG observations of molecular hydrogen in the CO line 
(3~mm, $J=1\to0$) \citep{helfer2003} taken from the {\sc ned} database and 
atomic hydrogen in the 21-cm line \citep{walter2008} from the THINGS 
project database,\footnote{http://www.mpia.de/THINGS} we estimated 
the characteristic wavelength $\Lambda$ of perturbations in the gaseous 
disk of the well-studied spiral galaxy NGC~628 (Fig.~\ref{figure:fig4}), 
which exhibits a distinct regularity in the distribution of the young 
stellar population along its spiral arms \citep{gusev2013}. It should 
be noted that we previously used these two-dimensional images of the 
galaxy in the CO and 21~cm lines in \cite{gusev2013} and 
\cite{gusev2015}, respectively.

The angular resolution for both HI and H$_2$ observations is 
$\approx6\arcsec$ (200~pc) \citep{helfer2003,walter2008}. In addition 
to HI intensity (surface density $\Sigma$) maps, the THINGS data also 
include radial velocity $v_{\rm rad}$ and gas velocity dispersion 
$\sigma{\rm (HI)}$ maps. The HI velocity dispersion map was employed in 
this study.

It should be noted that H$_2$ observations in NGC~628 were also carried 
out as part of the HERACLES \citep{leroy2009} and CARMA \citep{rahman2012} 
projects. The HERACLES data show slightly better sensitivity, but have 
poorer resolution than the BIMA SONG data \citep{marchuk2018}, while the 
CARMA data do not cover the outer regions of NGC~628. In general, the 
H$_2$ observations obtained in these three projects are in good agreement 
with each other \citep{leroy2009,marchuk2018}, except for the central 
region of the galaxy ($r<50\arcsec$), which is not addressed in this paper.

The conversion coefficients for transforming the fluxes in the CO and 
21-cm lines into H$_2$ and HI surface densities were taken from 
\cite{leroy2008}.

\begin{figure*}
\vspace{8mm}
\centerline{\includegraphics[width=17cm]{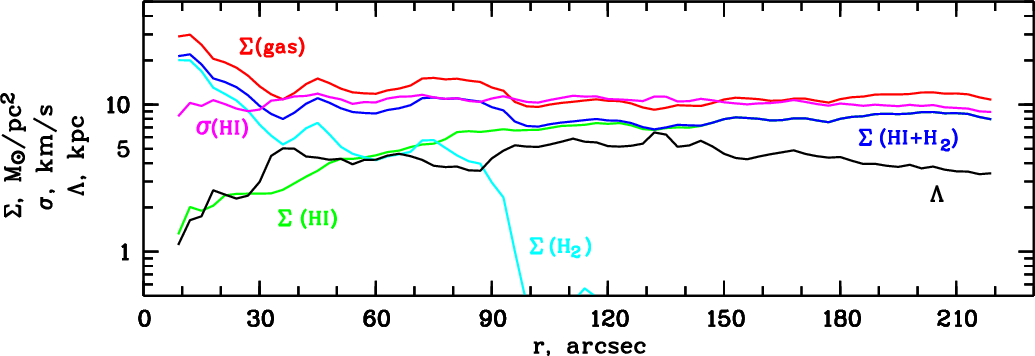}}
\caption{Radial profiles of surface densities $\Sigma{\rm (HI)}$ (green 
curve), $\Sigma{\rm (H_2)}$ (light blue), $\Sigma{\rm (HI+H_2)}$ (dark 
blue), $\Sigma{\rm (gas)}$ (red), velocity dispersion $\sigma{\rm (HI)}$ 
(purple), and characteristic star formation wavelength $\Lambda$ (black 
curve) in the galaxy NGC~628.
\label{figure:fig5}}
\end{figure*}

The radial distributions $\Sigma{\rm (HI)}$, $\Sigma{\rm (H_2)}$, and 
$\sigma{\rm (HI)}$ obtained by us, as presented in Fig.~\ref{figure:fig5}, 
align within the margin of error with similar distributions reported 
\citep{leroy2008,marchuk2018}. The only exceptions are the results on 
$\Sigma{\rm (HI)}$: based on the same data, \cite{leroy2008} and 
\cite{marchuk2018} obtained $\Sigma{\rm (HI)}$ values that differ by 
a factor of $\sim2$ (see the discussion in \cite{marchuk2018}). In this 
study, we decided to use the calibration provided by the THINGS project 
team from \cite{leroy2008}. It is important to note that the differences 
in $\Sigma{\rm (HI)}$ estimates will not significantly impact our results: 
atomic hydrogen is distributed more uniformly across the galaxy disk than 
molecular hydrogen, which is concentrated strictly within the spiral arms 
of NGC~628 (see \cite{leroy2008}, Fig.~35; \cite{helfer2003}, Fig.~6; 
\cite{gusev2013}, Fig.~14).

Additionally, we used optical images of NGC~628 obtained in our previous 
work (see \cite{gusev2013} and references therein).

Given that the galaxy is tilted toward us at an angle of $7\degr$, the 
projection effects on both surface gas densities and geometric parameters 
(coordinates, galactocentric distance) introduce a correction of less 
than $1\%$. Hence, in this study, we did not adjust the maps of NGC~628 or 
the $\Sigma{\rm (HI)}$ and $\Sigma{\rm (H_2)}$ values for the disk 
inclination.

Assuming that atomic and molecular hydrogen are well mixed (see, e.g., 
\cite{mogotsi2016}), we assume that their velocity dispersions are equal, 
$\sigma{\rm (HI)}=\sigma{\rm (H_2)}$, and that the speed of sound and 
velocity dispersion in the gas are equal: $c{\rm (gas)} = \sigma{\rm (gas)}$.

To account for the contribution of helium and heavier elements, 
a coefficient of 1.36 is used (see, e.g., \cite{marchuk2018}): 
$\Sigma{\rm (gas)} = 1.36(\Sigma{\rm (HI)}+\Sigma{\rm (H_2)})$.

Using the $\Sigma{\rm (HI)}$, $\Sigma{\rm (H_2)}$, and $\sigma{\rm (HI)}$ 
maps, we constructed radial profiles of surface densities and dispersion 
of gas velocities in the galaxy with annular apertures $3\arcsec$ wide 
(Fig.~\ref{figure:fig5}). The $3\arcsec$ step was selected to align 
THINGS data with the $1.5\arcsec$/pixel scale and the BIMA SONG data 
with $1\arcsec$/pixel scale.

The radial profile of the characteristic distance $\Lambda$ was calculated 
using formula~(\ref{equation:l1}), where $c_{\rm g}=\sigma{\rm (HI)}$ 
and $\Sigma_{\rm g}=1.36(\Sigma{\rm (HI)}+\Sigma{\rm (H_2)})$.

The resulting azimuthally averaged characteristic distances $\Lambda$ are 
several times larger than those expected from observations 
\citep{elmegreen1983,gusev2013}. However, it is worth noting that at 
distances $r=40-200\arcsec$ from the center of NGC~628, where regular 
chains of star-forming regions are observed, the $\Lambda$, 
$\Sigma_{\rm g}$, and $\sigma_{\rm g}$ values remain nearly constant and 
do not depend on $r$: $\Lambda=4.7\pm0.7$~kpc, 
$\Sigma_{\rm g}=12\pm2\,M_{\odot}$/pc$^2$, 
$\sigma_{\rm g}=10.7\pm0.5$~km/s. The HI surface density in the galaxy 
begins to decrease far beyond the development of spiral arms, at 
distances $r>300\arcsec$ from the center \citep{leroy2008}.

The approximate constancy of the parameters of surface density and gas 
velocity dispersion (speed of sound), which determine the characteristic 
star formation wavelength $\Lambda$ over a wide range of distances to 
the center of the galaxy, provides an explanation for the existence of 
regular chains of star formation regions in NGC 628's spiral arms. 
However, the average $\Lambda$ values obtained for $r=40-200\arcsec$ 
are $\approx3$~times larger than the observed distances in 
\cite{elmegreen1983} and $\approx12$~times larger than the estimates in 
\cite{gusev2013}.

\begin{figure*}
\centerline{\includegraphics[width=17cm]{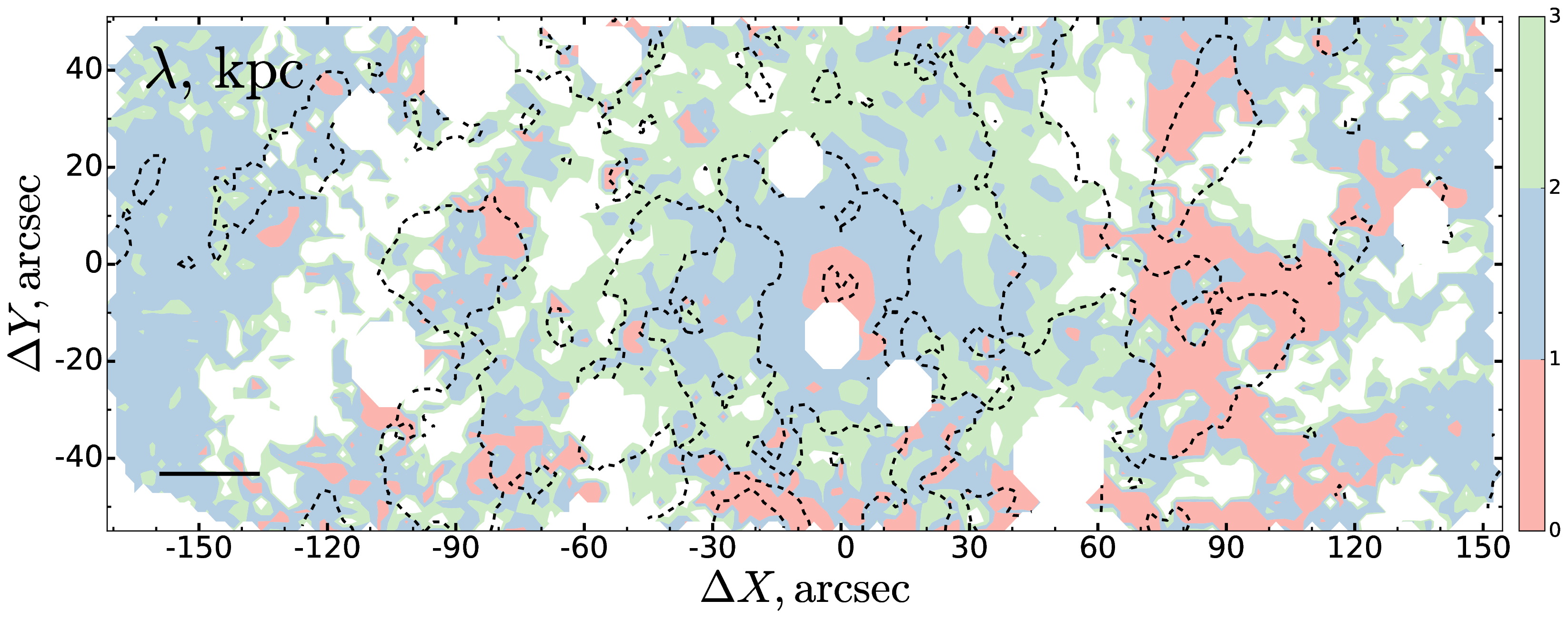}}
\caption{Most unstable wavelength $\lambda$ in kpc for the model of the 
NGC~628 stellar-gas disk. White areas show masked stars or regions 
$\lambda>3$~kpc. The black dotted lines represent regions of star 
formation. The black segment in the bottom left corner corresponds to 
a linear scale of 840~pc. North is up, east is to the left.
\label{figure:fig6}}
\end{figure*}

A map of the $\lambda$ values for NGC~628 was constructed using the 
$\Sigma{\rm (HI)}$, $\Sigma{\rm (H_2)}$, and $\sigma{\rm (HI)}$ maps. 
Figure~\ref{figure:fig4} depicts the isolines of $\Lambda=800$ and 
600~pc overlaid on the optical image of NGC~628.

As can be seen from Fig.~\ref{figure:fig4}, although the regions with 
minimum $\lambda$ values are concentrated along the spiral arms, the 
total area of the regions with $\Lambda<800$~pc is very small. The 
sizes of the largest individual regions do not exceed 700~pc.

According to the theory of gas disk instability, the averaging scale must 
be greater than the perturbation wavelength, i.e., the regions with 
$\Lambda<600-800$~pc should completely encompass the spiral arms of the 
galaxy and have a thickness of $\sim\Lambda$.

Thus, the direct determination of the $\Lambda$ values in the disk of 
NGC~628 obtained using formula~(\ref{equation:l1}) contradicts the results 
of \cite{gusev2013}, where a characteristic distance of 400~pc was found.

The stellar-gas disk of NGC~628 was analyzed by \cite{marchuk2018} based 
on data from the THINGS, HERACLES, and VENGA projects 
\citep{blanc2013a,blanc2013b}. The map of local $\lambda$ values for 
the stellar-gas disk of the galaxy is shown 
in Fig.~\ref{figure:fig6}.\footnote{Figure courtesy of A.A.~Marchuk 
(St.~Petersburg State University). It is similar but not identical to 
Fig.~4 from \cite{marchuk2018}.} As we can see, the values and 
distribution of parameter $\Lambda$ in the stellar-gas disk of NGC~628 
also do not agree with the results of \cite{gusev2013}: regions with 
$\lambda<1$~kpc occupy a negligible fraction of the total area, and their 
linear size does not exceed the inhomogeneity value $\Lambda$.

It should be noted that the calculations of spiral arm fragmentation 
performed in \cite{inoue2021} for NGC~628 also predict the characteristic 
distances $\Lambda>1$~kpc along the spirals.

\section{DISCUSSION OF THE RESULTS}

The characteristic regularity scales obtained through our method 
\citep{gusev2013,gusev2020} are several times smaller than those in 
\cite{elmegreen1983}. This can be explained by the fact that 
\cite{elmegreen1983} studied the distribution of large stellar complexes 
and HII~regions. Our technique allows us to investigate zones of enhanced 
star formation (the concentration of a young stellar population with an 
age of $t<300$~Myr). Stellar complexes of \cite{elmegreen1983} are only 
a specific case of such regions with a higher concentration of the young 
stellar population.

\begin{figure}
\centerline{\includegraphics[width=8.2cm]{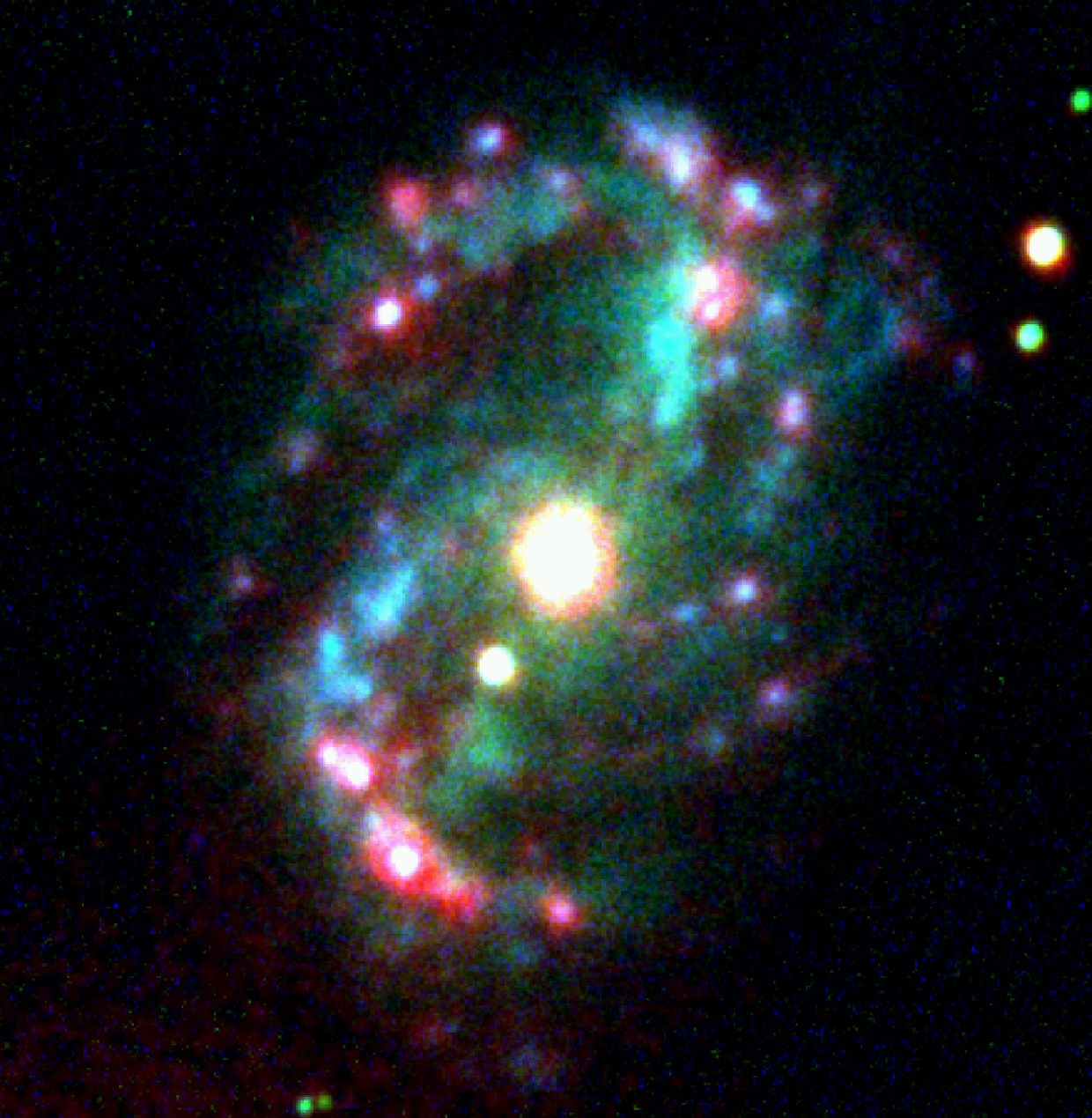}}
\caption{Composite image of NGC~6217 in $U$ (blue), $B$ (green), and 
H$\alpha$ (red). North is up, east is to the left.
\label{figure:fig7}}
\end{figure}

\begin{figure*}
\centerline{\includegraphics[width=16cm]{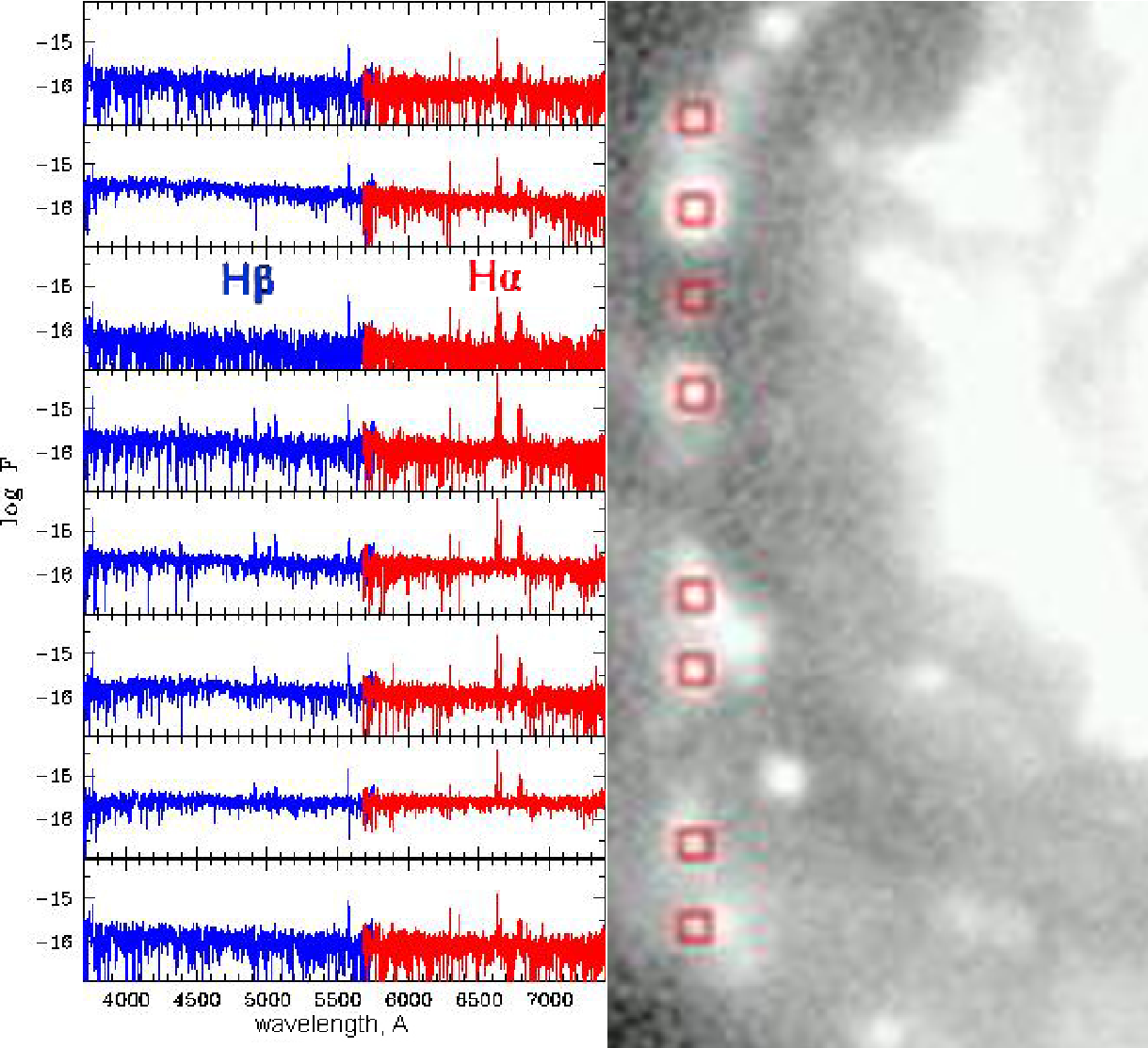}}
\caption{Southern spiral arm of the galaxy NGC~3963 (right) with a regular 
chain of star-formation regions (red squares) and spectra of the 
corresponding regions (left).
\label{figure:fig8}}
\end{figure*}

\cite{proshina2022} discovered a drift of star formation along the ring 
of NGC~4324, where relatively younger star formation regions often lie 
between relatively older ones. This regularity is not unique: younger and 
older regions alternate in most galaxies. NGC~6217 serves as an example 
(Fig.~\ref{figure:fig7}): its color image shows that the star formation 
regions comprising a regular chain have different ages. The youngest 
are the HII~regions, which are nearly invisible in broadband filters, 
while the oldest regions lack signs of emission in H$\alpha$ (see the 
morphological evolutionary sequence in \cite{whitmore2011}). Another 
example is the data of spectral observations from \cite{gusev2021} of 
a regular chain of star-forming regions in the spiral arm of the galaxy 
NGC~3963 (Fig.~\ref{figure:fig8}). Located at a distance of 
$\approx1.6$~kpc from each other, some regions have a pronounced emission 
spectrum (see the H$\beta$ line), while others do not.

In the case of synchronous onset of star formation in neighboring giant 
molecular clouds (GMCs), it is possible for an expanding shock wave from 
one star formation region, propagating along a spiral or ring, to 
encounter a shock wave from a neighboring source approximately at the 
midpoint.

The influence of the magnetic field on the regularity of star formation 
plays a secondary role. Unfortunately, studies of the magnetic field, 
even in nearby galaxies such as NGC~628 and NGC~6946, have been carried out 
with a linear resolution $>1$~kpc \citep{beck1989,heald2009,frick2000}, 
which is insufficient to study the detailed influence of the magnetic 
field on the scales of interest. In any case, the magnitude of the 
magnetic field has little effect on $\Lambda$; it can reduce $\Lambda$ 
by no more than $20\%$ (see Introduction).

It should be noted that the $\lambda$ values obtained from the analysis 
of the distribution histograms of distances between neighboring 
star-forming regions and the data from the Fourier analysis do not always 
agree. This is particularly noticeable in the study of NGC~6946. This 
discrepancy appears to be associated either with a quasi-regular 
distribution of star-forming regions (neighboring regions are located in 
a narrow but significant distance range), or with the presence of separate 
local regular chains of star-forming regions in a spiral arm or ring.

The results obtained on the example of NGC~628 show that the azimuthally 
averaged star formation wavelength $\Lambda$ can remain approximately 
constant over a wide range of galactocentric distances. On the one 
hand, the constancy of $\Lambda$ in a classical spiral galaxy like 
NGC~628 may indicate that the constancy of parameter $\Lambda$ and, 
consequently, the presence of regularity in the distribution of the 
young stellar population, could be a relatively common occurrence 
in galaxies. On the other hand, this provides theoretical support for 
the strict regularity in the distribution of star-forming regions in 
the NGC~628's spiral arms.

However, an unresolved contradiction remains. All theoretical models 
concerning the gravitational instability of the disk and fragmentation 
of spiral arms, filaments, and rings predict regularity with $\Lambda$ 
equal to several kiloparsecs, which is several times larger than what 
is observed. The mechanism behind the formation of regularities on 
scales of 350–500~pc is unknown and requires further research.

The question arises: can there be several scales of regularity on 
different spatial scales? The following results may suggest this 
possibility: (1) in NGC~628, several distinct characteristic distances 
are observed for different samples of objects, with brighter and larger 
stellar complexes being located at greater distances from each other 
\citep{gusev2013}; (2) in the nearest galaxy under study, NGC~6946 
($d=5.9$~Mpc), characteristic distances of $\sim300$~pc are found; 
(3) according to \cite{elmegreen1983,elmegreen2018}, the characteristic 
distances between neighboring star formation regions are approximately 
equal to three times the diameter ($D$) of these regions. In the case 
of NGC~4321, where \cite{elmegreen2018} studied IR sources smaller in 
size than the stellar complexes from \cite{elmegreen1983}, the relation 
$\Lambda\approx3D$ is still valid. It should be noted that in NGC~628 
the hierarchy of stellar concentrations is maintained with a consistent 
fractal dimension of 1.5 in the scale range from 2~pc to 1~kpc 
\citep{elmegreen2006,gusev2014}.

Further research is needed to address this issue.

The minimum characteristic distances $\Lambda$ obtained are 5–10 times 
larger than the linear resolution of the images of the corresponding 
galaxies. Only in the case of the S1 spiral arm of the NGC~895 galaxy, 
Fourier analysis revealed a $\Lambda$ scale three times greater than 
the linear resolution of NGC~895 images. At the same time, there is 
no direct relationship between the linear image resolution and the 
characteristic distance $\Lambda$. As mentioned earlier, the $\Lambda$ 
value apparently depends on the diameter of the star-forming regions. 
On the other hand, the sizes of young stellar groups at different 
hierarchical levels (complexes, aggregates, associations, clusters 
\citep{efremov1987}) are approximately the same across different 
galaxies.

\section{CONCLUSIONS}

Recent studies of a number of galaxies (NGC~628, NGC~895, NGC~4321, 
NGC~4324, NGC~5474, NGC~6217, and NGC~6946) have shown that regularity 
in the spatial distribution of the young stellar population along 
the spiral arms and rings of galaxies is observed more often than 
expected. This phenomenon has been observed in galaxies of various 
morphologies from S0 to Scd.

In the majority of these galaxies, the characteristic distance between 
neighboring zones of concentration of the young stellar population is 
equal to or a multiple of 350–500~pc.

The presence or absence of shock waves does not affect the formation 
of regular chains of star formation regions along galactic spirals and 
rings.

\section*{ACKNOWLEDGMENTS}

I would like to thank the anonymous reviewer for the valuable comments. 
I express my gratitude to the organizers of the conference Modern 
Stellar Astronomy 2022. I thank E.V.~Shimanovskaya (SAI MSU), A.V.~Zasov 
(SAI MSU), O.V.~Egorov (SAI MSU, Heidelberg University), A.A.~Marchuk 
(St. Petersburg State University), and N.A.~Zaitseva (SAI MSU) for help 
and advice, as well as A.V.~Moiseev (SAO RAS) for the fruitful discussion. 
Special thanks to A.A.~Marchuk for the original $\lambda$ distribution 
map. This study used open data from the THINGS, BIMA SONG, SDSS projects 
and the {\sc ned} and {\sc leda} databases.

\section*{FUNDING}

This study was supported by the Russian Foundation for Basic Research 
(RFBR), grant no. 20-02-00080. The study was supported by the 
Interdisciplinary Scientific and Educational School of Moscow State 
University ''Fundamental and Applied Space Research.''

\section*{CONFLICT OF INTEREST}

The author declares that he has no conflicts of interest.

\vspace{4mm}

{\it The paper was translated by M.~Chubarova}

\end{document}